\documentclass[12pt,preprint]{aastex}
\usepackage{natbib}
\usepackage{graphicx}
\usepackage{slashbox}
\usepackage{multirow}
\usepackage{mathrsfs,amssymb}
\usepackage{amsmath}
\newcommand       \Angstrom     {\,{\rm \AA}}

\newcommand       \CtoH        {{\small \left[{\rm C/H}\right]}}

\newcommand       \Sstar       {{\small \left[{\rm S/H}\right]_\star}}

\newcommand       \rgd         {r_{\rm g/d}}
\newcommand       \Smgs        {\phi_{\rm MgS}^{S}}
\newcommand       \cm           {\,{\rm cm}}

\newcommand       \erg          {\,{\rm erg}}

\newcommand   \g        {\,{\rm g}}
\newcommand       \K            {\,{\rm K}}

\newcommand   \kpc      {\,{\rm kpc}}
\newcommand   \s        {\,{\rm s}}

\newcommand       \Qabs         {Q_{\rm abs}}
\newcommand       \Qabsuv   {Q_{\rm abs}^{\rm UV}}
\newcommand       \Qabsir   {Q_{\rm abs}^{\rm IR}}

\newcommand       \qabsuv       {Q_{\rm abs}^{\rm UV}}
\newcommand       \Qo           {Q_{\rm o}}
\newcommand       \lambdao      {\lambda_{\rm o}}
\newcommand       \simlt        {\lesssim}
\newcommand       \simgt        {\gtrsim}

\newcommand       \mum          {\,{\rm \mu m}}

\newcommand       \Teff         {T_{\rm eff}}

\newcommand       \mMgS         {M_{\rm MgS}}
\newcommand       \simali       {\sim\,}

\newcommand       \rstar        {r_\star}
\newcommand       \Fstar        {F_{\lambda}^{\star}}

\shorttitle{On MgS as the 30$\mum$ Feature Carrier}
\title{
 \vspace*{1.0em}
On Magnesium Sulfide as the Carrier of the 30$\mum$ Emission Feature
in Evolved Stars
     }

\author{Ke Zhang\altaffilmark{1}, B.W. Jiang\altaffilmark{1,2}
                                 and Aigen Li\altaffilmark{2}}
\altaffiltext{1} {Department of Astronomy, Beijing Normal University,
                  Beijing 100875, China; {\sf zhangke@mail.bnu.edu.cn,
                  bjiang@bnu.edu.cn}}
\altaffiltext{2} {Department of Physics and Astronomy,
                  University of Missouri,
                  Columbia, MO 65211, USA;
                  {\sf lia@missouri.edu}}

\begin{document}

\begin{abstract}
A large number of carbon-rich evolved objects (asymptotic giant
branch stars, protoplanetary nebulae, and planetary nebulae) in both
the Milky Way galaxy and the Magellanic Clouds exhibit an enigmatic
broad emission feature at $\simali$30$\mum$. This feature, extending
from $\simali$24$\mum$ to $\simali$45$\mum$, is very strong and
accounts for up to $\simali$30\% of the total infrared luminosity of
the object. In literature it is tentatively attributed to magnesium
sulfide (MgS) dust. Using the prototypical protoplanetary nebula
around HD\,56126 for illustrative purpose, however, in this work
we show that in order for MgS to be responsible for
the 30$\mum$ feature, one would require an amount of MgS mass
substantially exceeding what would be available
in this source. We therefore argue that MgS is unlikely the
carrier of the 30$\mum$ feature seen in this source
and in other sources as well.
\end{abstract}

\section{Introduction\label{sec:intro}}
In carbon-rich evolved objects, there are two prominent, mysterious
emission features known as the ``21$\mum$'' and ``30$\mum$''
features (e.g. see \citealt{Jiang2009EPS}). The
21$\mum$ feature is seen almost exclusively in 16 protoplanetary
nebulae (PPNe) and its identification is notoriously difficult (see
\citealt{Posch2004ApJ, Zhang2009MNRAS.tmp..663Z}). The 30$\mum$
feature, first discovered by \citet{Forrest1981ApJ...248..195F} in
the {\it Kuiper Airborne Observatory} spectrometry of C stars and
planetary nebulae (PNe), is very broad and strong,
extending from $\simali$24$\mum$ to $\simali$45$\mum$ and accounting
for up to $\simali$30\% of the total infrared (IR) luminosity of the
object
\citep{Volk2002ApJ...567..412V}.\footnote{%
  The ISO spectroscopy suggested that the 30$\mum$ feature
  consists of two subfeatures: a narrow feature at 26$\mum$
  and a broad one at 33$\mum$
  \citep{Hrivnak2000ApJ...535..275H, Volk2002ApJ...567..412V}.
  But this was not confirmed by {\it Spitzer}
  \citep{Hrivnak2009ApJ}.
  }
More ubiquitously seen in C-rich objects
than the 21$\mum$ feature,
the 30$\mum$ feature has been detected
in 63 Galactic and 25 Magellanic objects
(see \citealt{Jiang2009EPS} and references therein),
including asymptotic giant branch (AGB) stars,
post-AGB stars and PNe.\footnote{%
  The 21$\mum$ sources all exhibit the 30$\mum$ feature.
  But these two features do not appear to correlate
  in their strengths (see \citealt{Jiang2009EPS}).
  Unlike the 21$\mum$ feature which displays little
  shape variation,
  the 30$\mum$ feature varies in its peak wavelength
  and width among different sources
  (e.g. see \citealt{Hrivnak2000ApJ...535..275H,
                     Hony2002A&A...390..533H}).
  }

Compared to the 21$\mum$ feature for which over a dozen carrier
candidates have been proposed, the identification of the 30$\mum$
feature has not received the attention it deserves.
Magnesium sulfide (MgS) solids were first proposed by
\citet{Goebel1985ApJ...290L..35G} as a carrier of
the 30$\mum$ feature, based on (1) the similarity of the observed
emission spectral profiles and the laboratory spectra of MgS
\citep{Nuth1985ApJ...290L}, and (2) the considerations
of C-rich equilibrium condensation chemistry which predicts
the condensation of MgS in C-rich stars \citep{Lattimer1978ApJ}.
The MgS proposition gains further support from detailed modeling
of the 30$\mum$ feature of a large number of C-rich sources
obtained by the {\it Infrared Space Observatory} (ISO)
\citep{Jiang1999A&A...344..918J, Szczerba1999A&A,
Hony2002A&A...390..533H}.

A valid carrier candidate should not only reproduce the observed
emission spectra but also satisfy the abundance constraints (i.e.,
the amounts of elements required to lock up in the carrier should
not exceed what are available in the 30$\mum$ sources). The
absorption profile $Q_{\rm abs}(\lambda)$ of MgS dust with a
distribution of ellipsoidal shapes exhibits a broad band around
30$\mum$. To fit the observed 30$\mum$ emission spectra, one often
compares the observed spectra with $Q_{\rm
abs}(\lambda)\,B_\lambda(T)$, the product of the absorption
efficiency and a blackbody with an assumed temperature $T$
(e.g. see \citealt{Hony2002A&A...390..533H}).

The MgS temperature is an essential parameter:
it not only affects the shape and peak wavelength
of the model emission spectrum but also determines
the total amount of MgS dust $\mMgS$ required
to account for the flux $F_\lambda$ emitted from the
30$\mum$ band: $\mMgS \propto
F_\lambda/\left[Q_{\rm abs}(\lambda)\,B_\lambda(T)\right]$.
Previous models based on MgS do not seem to have a sulfur (S)
budget problem \citep{Jiang1999A&A...344..918J, Szczerba1999A&A,
Hony...2003A&A...402..211H}. However, the MgS temperatures adopted
in these models were often treated as a free parameter and were
derived from matching the spectral profiles of the observed
30$\mum$ feature with $Q_{\rm abs}(\lambda)\,B_\lambda(T)$
by varying $T$ (e.g. see \citealt{Hony2002A&A...390..533H}).
Due to the lack of the dielectric functions
in the ultraviolet (UV), visible and near-IR wavelength ranges
for MgS dust, it is not possible to calculate the precise
thermal equilibrium temperatures of MgS dust in the 30$\mum$
sources. Therefore, the previously derived S budget may be
questionable.

In this work we make a rather generous estimation of the UV/visible
absorptivity of MgS and calculate its temperatures in the
circumstellar shell of HD\,56126, a prototypical source of the
21$\mum$ and 30$\mum$ features. It is shown that the required MgS
dust mass exceeds what would be available by a substantial amount.
We therefore argue that MgS is unlikely the carrier of the 30$\mum$
feature.

\section{Optical Properties of MgS Dust
         \label{sec:optical properities}}
In order to calculate the temperature of a dust
species, the knowledge of its optical properties over a wide
wavelength range is required. Unfortunately, the dielectric
functions of MgS have only been experimentally determined in the IR
(e.g. see \citealt{Begemann1994ApJ...423L..71B,
Hofmeister2003MNRAS}). The UV/visible absorptivities of MgS dust are
essential since they determine how much energy will be absorbed by a
given MgS grain (when exposed to starlight) which in turn determines
how much energy will be emitted in the IR. It is this balance of
energy between absorption and emission that determines the grain
temperature.

\begin{figure}
  \includegraphics[width=12cm,angle=90]{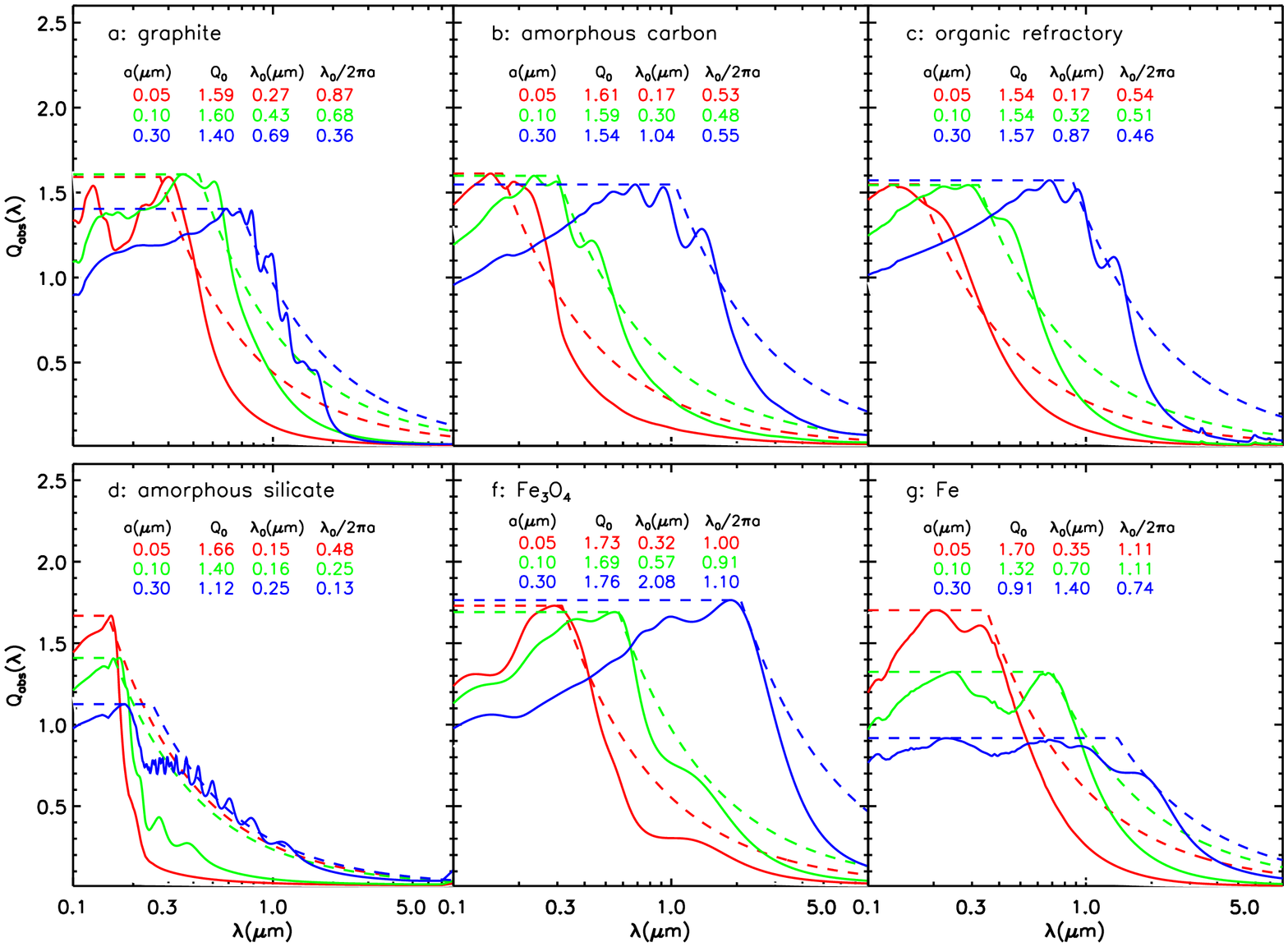}\\[8mm]
  \caption{\footnotesize
           Absorption efficiencies $Q_{\rm abs}(\lambda)$
           of both dielectric (amorphous carbon,
           organic refractory, amorphous silicate) dust
           and metallic dust (graphite, iron, magnetite)
           calculated from Mie theory for three sizes:
           $a$\,=\,0.05 (red), 0.1 (green), 0.3$\mum$ (blue).
           Also shown are the best fits (dashed lines) given
           by eq.1 (with $Q_{\rm o}$ and $\lambda_{\rm o}$
           treated as free parameters).
           }
  \label{fig:1}
\end{figure}

We take the following procedure to approach the UV/visible
absorption efficiency $Q_{\rm abs}^{\rm UV}(\lambda)$
of MgS dust:
\begin{equation}\label{eq:qabs_UV}
\qabsuv(\lambda) = \left\{\begin{array}{lr}
Q_{\rm o} ~, ~~~~~~~~\lambda \le \lambda_{\rm o} ~;\\
Q_{\rm o}\,\left(\lambda_{\rm o}/\lambda\right) ~,
          ~~ \lambda > \lambda_{\rm o} ~;\\
\end{array}\right.
\end{equation}
where $\Qo$ is a constant and the cut-off wavelength
$\lambdao$ depends on grain size (and material).
What would be the reasonable $\Qo$ and $\lambdao$ values
for MgS? To shed light on this, we use Mie theory to
calculate the UV to near-IR absorption efficiencies for
spherical grains of various compositions and sizes.
We consider both dielectric
(amorphous silicate, amorphous carbon, organic refractory)
and metallic materials (graphite, iron Fe, magnetite Fe$_3$O$_4$).
We then approximate the calculated absorption profiles
$Q_{\rm abs}(\lambda)$ with eq.\ref{eq:qabs_UV}.
For submicron-sized dust, as shown in Figure\,\ref{fig:1},
while $\Qo\approx 1.6$ is a reasonable estimation for
both dielectric and metallic dust, the cut-off wavelength
for a given size $a$ is generally shorter for dielectric
dust ($\lambdao\approx \pi a$)\footnote{%
  For amorphous silicate dust larger than $\simali$0.07$\mum$
  in size, $\lambdao$ is much smaller than $\pi a$.
  This is due to the sudden rise in absorption caused by
  the electronic transition at $\lambda$\,$\simali$0.1--0.2$\mum$
  \citep{Kim1995ApJ...442..172K}.
  }
than that for metallic dust ($\lambdao\approx 2\pi a$).
For MgS dust of size $a$,
we thus adopt $\Qo = 1.6$ and $\lambdao = \pi a$.
We note that, as can be seen in Figure\,\ref{fig:1},
eq.\ref{eq:qabs_UV} with $\Qo = 1.6$ and $\lambdao = \pi a$
(or $\lambdao = 2\pi a$) almost always overestimates
the actual UV/visible absorptivities of all six dust species,
particularly at $\lambda <\lambdao$.
This suggests that the MgS temperatures derived using
the UV/visible absorptivity of eq.\ref{eq:qabs_UV} would
be overestimated and thus one would underestimate $\mMgS$
(since it is most likely that eq.\ref{eq:qabs_UV} also overestimates
the actual UV/visible absorptivity of MgS).

\begin{figure}
  \includegraphics[width=12cm]{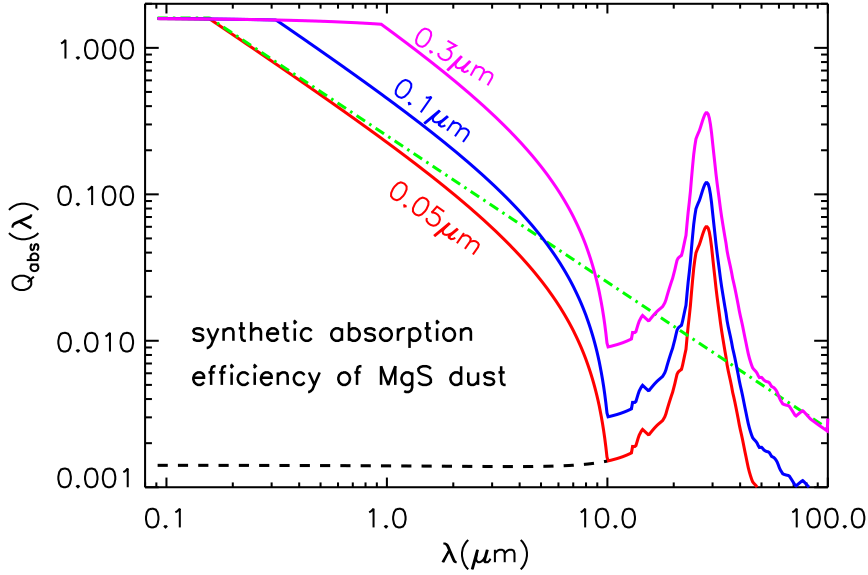}\\[8mm]
  \caption{\footnotesize
           Synthetic absorption efficiencies of MgS dust
           of $a$\,=\,0.05 (red), 0.1 (blue), 0.3$\mum$ (magenta).
           For $a$\,=\,0.05$\mum$ we also show
           the UV/visible component $\Qabsuv$ (green dot-dashed line)
           and the IR component $\Qabsir$ (black dashed line) alone.
           }
  \label{fig:synthetic Qabs}
\end{figure}

In the IR, we will use the dielectric functions of
Mg$_{0.9}$Fe$_{0.1}$S measured by
\citet{Begemann1994ApJ...423L..71B}
in the 10--200$\mum$ wavelength range to calculate
the absorption efficiencies $\Qabsir(\lambda)$ of MgS.
At longer wavelengths we will extrapolate assuming
$\Qabsir(\lambda)\propto \lambda^{-2}$.
We will consider both spherical dust and
dust with a CDE (continuous distribution of ellipsoids)
shape distribution \citep{Bohren1983asls.book}.
%
Finally, we smoothly join $\Qabsuv(\lambda)$
and $\Qabsir(\lambda)$ through
\begin{equation}\label{eq:UV+IR}
\Qabs(\lambda) = (1-\xi_{\rm IR})\, \Qabsuv(\lambda)
               + \xi_{\rm IR} \Qabsir(\lambda) ,
              ~~912\,\Angstrom <\lambda <1\,{\rm cm}~ \\
\end{equation}
\begin{equation}\label{eq:xiIR}
\xi_{\rm IR}(\lambda) =
\min\left[1, \left(\lambda/10\mu{\rm m}\right)^3\right] ~.
\end{equation}
The absorption efficiency synthesized from eq.\ref{eq:UV+IR}
is dominated by $\Qabsuv(\lambda)$ at $\lambda<5\mum$
and by $\Qabsir(\lambda)$ at $\lambda>10\mum$
(see Figure\,\ref{fig:synthetic Qabs}).

\section{The Tester: HD 56126}\label{sec:HD 56126}
A successful candidate carrier should be able to explain the
observed 30$\mum$ feature in all sources. A failure in a single
source would be sufficient to rule out the candidate. To examine
whether the carriers can account for the observed feature strength,
we choose HD\,56126, a prototypical source of
the 21$\mum$ and 30$\mum$ features and one of
the best studied post-AGB stars, as the tester.

\citet{VanWinckel2000A&A...354} performed a homogeneous photospheric
abundance analysis for HD\,56126, and derived the abundances of
sulfur and carbon to be ${\rm S/H \approx 4.07\times 10^{-6}}$, and
${\rm C/H \approx 4.47\times 10^{-4}}$. HD\,56126 has a radius of
$\rstar\approx 49.2\,r_\odot$ ($r_\odot$ is the solar radius), and a
luminosity of $L_\star\approx 6054\,L_\odot$ ($L_\odot$ is the solar
luminosity). We approximate the stellar radiation by the Kurucz
model atmospheric spectrum with $\Teff = 7250\K$ and $\log g=1.0$.
We adopt a distance of $d\approx 2.4\kpc$ from Earth to HD\,56126.

The size and mass of the circumstellar envelope of
HD\,56126, which are important for determining the quantity
of MgS dust available in the envelope, are still controversial.
\citet{Hony...2003A&A...402..211H} suggested that the dust is
confined to the envelope with radius between
1.2$^{\prime\prime}$--2.6$^{\prime\prime}$
based on their mid-IR imaging at 11.9$\mum$.
Their detailed radiative transfer model
derived a circumstellar envelope mass of
$M_{\rm H}$\,$\sim$\,0.16--0.44\,${\rm M_\odot}$,
depending on the assumed gas-to-dust ratio ($\simali$220--600).
\citet{Meixner2004ApJ...614..371M} found a more
extended envelope (with radius between
1.2$^{\prime\prime}$--7$^{\prime\prime}$)
based on their CO J\,=\,1--0 line emission images.
They derived a much smaller mass for the envelope
($M_{\rm H}\sim$\,0.059\,${\rm M_\odot}$).

\section{How Much MgS Dust Is Required?}\label{sec:req}
\begin{figure}
  \includegraphics[width=12cm]{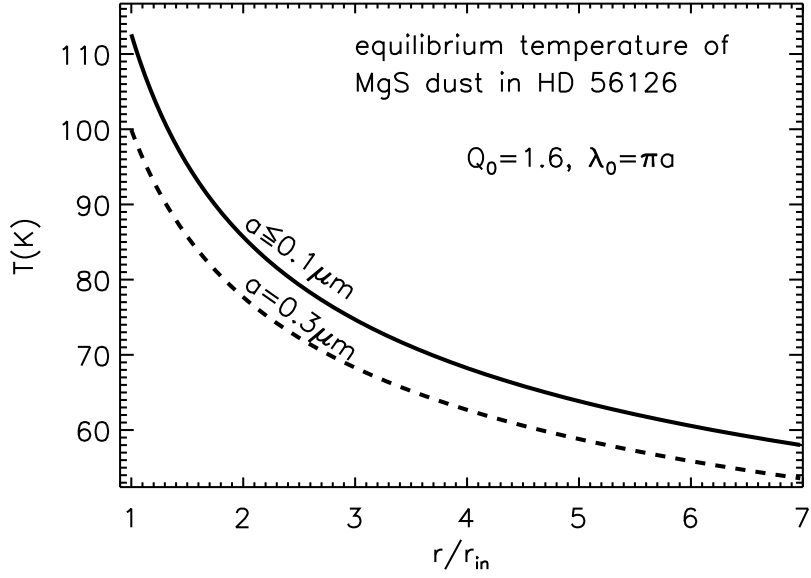}\\[8mm]
  \caption{\footnotesize
           Equilibrium temperatures of MgS dust
           in the HD\,56126 envelope.
           For dust with $a\simlt 0.1\mum$,
           $T$ is insensitive to $a$
           (see Footnote-\ref{ftnt:TvsA}).
           }
  \label{fig:Teq}
\end{figure}

By assuming the envelope is optically thin,\footnote{%
  If the envelope is not optically thin
  the conclusion of this work would be strengthened
  since when exposed to an attenuated starlight radiation,
  the MgS model would require more MgS dust to account for
  the same 30$\mum$ feature strength.
  }
we calculate the temperature $T(r,a)$
of MgS dust of size $a$ at a distance $r$ from the central star
from the energy balance between absorption and emission
\begin{equation}\label{eq:Teq}
\left(\frac{r_\star}{2r}\right)^2 \int^{\infty}_{0}\pi a^2 Q_{\rm
abs}(a,\lambda) \Fstar d\lambda = \int^{\infty}_{0}\pi a^2 Q_{\rm
abs}(a,\lambda) 4\pi B_\lambda\left(T[r,a]\right)d\lambda ~~,
\end{equation}
where $\Fstar$ is the stellar atmospheric flux
which is approximated by a Kurucz model.
In Figure\,\ref{fig:Teq} we show the equilibrium temperatures
of MgS dust of $a=0.1, 0.3\mum$ as a function of distance from
the central star.\footnote{%
  \label{ftnt:TvsA}
  For dust in the size range of $0.01< a< 0.1\mum$,
  $T$ is insensitive to $a$ since both $\Qabsuv/a$ and
  $\Qabsir/a$ are independent of $a$.
  This is because for $\lambda>0.3\mum$ where there is
  significant stellar radiation, $\Qabsuv \propto \lambdao \propto a$,
  while in the IR the dust is in the Rayleigh regime and therefore
   $\Qabsir \propto a$.
  }
We see that for $a\simlt 0.1\mum$ $T$ decreases from
$\simali$110\,K at the inner most shell to $\simali$50\,K
at the outer boundary, significantly lower than that
adopted by \citet{Hony2002A&A...390..533H}, $T=150\K$.
Larger dust has a lower $T$ (e.g. for dust of $a=0.3\mum$,
$T$ is lower than that of $a=0.1\mum$ by about $\simali$10\,K).

The IR emission per unit mass of MgS dust of size $a$ is
\begin{equation}\label{eq:Fmgs}
F_{\rm MgS}(\lambda)=\int^{r_{\rm out}}_{r_{\rm in}}
\left[3\,Q_{\rm abs}(a,\lambda)/4a\rho_{\rm MgS}\right]
4\pi B_\lambda\left(T[r,a]\right)\,dn(r)/dr\,4\pi r^2\,dr ~~,
\end{equation}
where $\rho_{\rm MgS}\approx 2.84\g\cm^{-3}$ is the mass density
of MgS dust, $r_{\rm in}\approx 1.2^{\prime\prime}$
and $r_{\rm out}\approx 7^{\prime\prime}$
are respectively the inner and outer radius
of the dust shell of HD\,56126
\citep{Meixner2004ApJ...614..371M},
and $dn(r)/dr$ is the MgS dust spatial density distribution.
We will consider two spatial distribution functions:
(1) $dn/dr\varpropto r^{-2}$, which means the mass loss of
the central star is constant during the whole process;\footnote{%
   The much more complicated function adopted by
   \citet{Meixner2004ApJ...614..371M}
   to reproduce the observed dust IR emission spectral energy
   distribution (SED) and mid-IR images of HD\,56126
   is actually similar to $dn/dr\varpropto r^{-2}$.
   }
and (2) $dn/dr\varpropto r^{-1}$, which was shown by
\citet{Hony...2003A&A...402..211H} to closely fit
the observed SED and mid-IR images (with the dust
confined to a narrow zone of
1.2$^{\prime\prime}$--2.6$^{\prime\prime}$ from the star).

The power output per unit mass in the 30$\mum$ band is
calculated by integrating $F_{\rm MgS}(\lambda)$
over the entire band (with the continuum subtracted)
\begin{equation}\label{eq:Etot}
E_{\rm MgS}^{\rm 30\mu{\rm m}} = \int_{30\mu{\rm m}\,{\rm band}}
         F_{\rm MgS}(\lambda)\,d\lambda ~~.
\end{equation}
The MgS mass $M_{\rm MgS}^{\rm req}$ required to account for the
observed 30$\mum$ emission in HD\,56126 is
\begin{equation}\label{eq:mmgs}
M_{\rm MgS}^{\rm req} =
E_{30\mu{\rm m}}^{\rm obs}/E_{\rm MgS}^{30\mu{\rm m}} ~~,
\end{equation}
where $E_{30\mu{\rm m}}^{\rm obs}
\approx$\,2$\times$10$^{36}$\,erg\,s$^{-1}$
is the total power emitted from the 30$\mum$ feature
of HD\,56126 \citep{Hony...2003A&A...402..211H}.

We calculate $M_{\rm MgS}^{\rm req}$ and tabulate
the results in Table\,\ref{tab:mgsmass_1}.
We see that for dust with a size $a\simgt 0.1\mum$,
$M_{\rm MgS}^{\rm req}$ increases with $a$ because
the dust temperature $T$ decreases with $a$
(see Figure\,\ref{fig:Teq}).
But for $a<0.1\mum$, $M_{\rm MgS}^{\rm req}$ is
insensitive to $a$ (since $T$ is insensitive to $a$
as long as $a$ is not in the nano size range and
the dust will not undergo stochastical heating by
single photons).
The required MgS mass is also affected by the adopted
dust spatial density distribution:
the $dn/dr \varpropto r^{-1}$ distribution requires
more MgS (since it places more MgS grains
in the cool, outer envelope region).
But in general, $M_{\rm MgS}^{\rm req}$ is in the order
of several tens of 10$^{-5}$\,$M_{\odot}$,\footnote{%
  This also appears unreasonably too high
  in view that the total dust mass in the envelope
  is just $M_{\rm dust}\approx 7.4\times 10^{-4}\,M_\odot$
  \citep{Hony...2003A&A...402..211H,
         Meixner2004ApJ...614..371M}.
  }
exceeding what would be available in HD\,56126 roughly by one order
of magnitude (see \S\ref{sec:ava}).

The 30$\mum$ feature arising from spherical MgS dust is too sharp to
be comparable with the 30$\mum$ feature seen in C-rich objects. This
is why a CDE shape distribution is often invoked (e.g. see
\citealt{Hony2002A&A...390..533H}). Unfortunately, the simple
formula of \citet{Bohren1983asls.book} for calculating the
absorption cross sections of CDE dust is only valid in the Rayleigh
regime (i.e. $a\ll \lambda$). There is no precise way to derive the
UV/visual absorption efficiencies of CDE dust (which are crucial to
determine its heating rate and equilibrium temperature). We
therefore simply adopt the equilibrium temperatures calculated for
its spherical counterpart. The MgS mass required to account for the
observed 30$\mum$ emission power is derived from eqs.\ref{eq:Fmgs},
\ref{eq:Etot} and \ref{eq:mmgs} with $\Qabsir$ calculated from the
CDE shape distribution (\citealt{Bohren1983asls.book}). As shown in
Table\,\ref{tab:mgsmass_1}, the CDE model requires $\simali$30\%
less MgS mass than the spherical model, but still needs much more
than what could be available in HD\,56126.

\section{How Much MgS Dust Would Be Available in HD\,56126?}
        \label{sec:ava}
Let $M_{\rm MgS}^{\rm ava}$ be the MgS dust mass available in the
circumstellar envelope of HD\,56126. Let $M_{\rm env}$, $M_{\rm H}$,
and $M_{\rm dust}$ be the total gas, hydrogen and dust mass in the
envelope. If we know the sulfur abundance $\Sstar$, the fraction of
sulfur depleted in MgS $\Smgs$, and the gas-to-dust ratio $\rgd$ of
HD\,56126, we can estimate $M_{\rm MgS}^{\rm ava}$ from
\begin{equation}\label{eq:Mmgs}
M_{\rm MgS}^{\rm ava} = \mu_{\rm MgS}\,M_{\rm H}\,\Sstar\,\Smgs ~~,~~
M_{\rm H} = M_{\rm env}/1.4 ~~,~~ M_{\rm env} = \rgd M_{\rm dust} ~~,
\end{equation}
where $\mu_{\rm MgS}$ is the molecular weight of MgS,
and the factor of 1.4 accounts for He whose abundance is
$\simali$10\% of H.

While both \citet{Hony...2003A&A...402..211H}
and \citet{Meixner2004ApJ...614..371M} derived
the circumstellar dust mass to be
$M_{\rm dust}\approx 7.4 \times10^{-4}\,M_\odot$,
they adopted a very different gas-to-dust ratio $\rgd$.
Assuming that all C atoms are locked up in CO,
\citet{Meixner2004ApJ...614..371M} estimated
$M_{\rm env}\approx 0.059\,M_\odot$ from the millimeter
interferometry images of the CO J\,=\,1--0 line.
This corresponds to a gas-to-dust ratio of $\rgd \approx 75$,
lower by a factor of $\simali$3 than the typical gas-to-dust
ratio of $\rgd \sim 200$ for carbon stars
\citep{Jura1986ApJ...303..327J}.
\citet{Hony...2003A&A...402..211H} argued for
a much higher gas-to-dust ratio, with $\rgd$\,$\simali$220--600.

If we assume that all S atoms are tied up in MgS
(i.e. $\Smgs=1$), with $\Sstar \approx 4.07\times 10^{-6}$
\citep{VanWinckel2000A&A...354}
and $M_{\rm dust}\approx 7.4 \times10^{-4}\,M_\odot$,
the total amount of MgS mass available in HD\,56126 would be
$M_{\rm MgS}^{\rm ava}$\,$\approx$\,
$9.60\times 10^{-6}$,
$2.60\times 10^{-5}$,
and $7.16\times 10^{-5}$\,$M_\odot$
respectively for $\rgd$\,=\,75, 220, and 600
(see Table\,\ref{tab:mgsmass_2}).
Even with $\rgd$\,=\,600,
$M_{\rm MgS}^{\rm req}$
appreciably exceeds what is available in HD\,56126!
Therefore, MgS is unlikely the carrier of the 30$\mum$ feature.

\section{Discussion}\label{sec:discussion}
While the gas-to-dust ratio of $\rgd=600$ adopted by
\citet{Hony...2003A&A...402..211H} may be in the high end,
$\rgd\approx 75$ \citep{Meixner2004ApJ...614..371M}
is probably too low.\footnote{%
  With $\rgd\approx 75$
  (corresponding to $M_{\rm env} \approx 0.059\,M_\odot$),
  the star (whose mass is $\simali$0.6\,$M_\odot$) would
  just have a ZAMS mass of $M_\star \approx 0.66\,M_\odot$.
  Such a low-mass star, even with a metallicity [Fe/H]\,=\,$-1$
  can not evolve to the tip of RGB in the cosmic age
  \citep{demarque2008arXiv0801.0451D},
  not to mention the post-AGB phase.
  }
The problem probably lies in the fact that
\citet{Meixner2004ApJ...614..371M} could have underestimated the
total gas mass $M_{\rm env}$ by assuming that all carbon atoms are
in CO. As a matter of fact, a considerable fraction of the C atoms
are in atomic carbon (CI): \citet{Knapp2000ApJ...534..324K} observed
the 609$\mum$ $^{3}P_1$\,$\rightarrow$\,$^{3}P_0$ line of CI in the
envelope of HD\,56126 and estimated the CI to CO abundance ratio to
be [CI/CO]\,$\approx$\,0.4. In addition, a small fraction of the C
atoms should be C$_2$, C$_3$, and CN
(\citealt{Bakker1996A&A...310..893B, Hrivnak1999ApJ}).

Let us assume that CO, CI, and amorphous carbon dust
are the major sinks of C (i.e. we neglect C$_2$, C$_3$,
CN and PAHs). The C budget in the envelope of HD\,56126 is
\begin{equation}\label{eq:cratio}
\CtoH_\star = \CtoH_{\rm CO} + \CtoH_{\rm CI} + \CtoH_{\rm AC} ~~,
\end{equation}
where $\CtoH_\star \approx 4.47\times 10^{-4}$ is the stellar C
abundance of HD\,56126 \citep{VanWinckel2000A&A...354}, $\CtoH_{\rm
CO} = M_{\rm CO}/\left[\mu_{\rm CO} M_{\rm H}\right]$, $\CtoH_{\rm
CI}\approx 0.4\,\CtoH_{\rm CO}$ \citep{Knapp2000ApJ...534..324K},
and $\CtoH_{\rm AC} = M_{\rm AC}/\left[\mu_{\rm C} M_{\rm
H}\right]$, are respectively the C abundances tied up in CO, CI and
amorphous carbon, where $M_{\rm CO} \approx 5.54\times
10^{-4}\,M_\odot$ \citep{Meixner2004ApJ...614..371M} and $M_{\rm AC}
\approx 7.4\times 10^{-4}\,M_\odot$
\citep{Hony...2003A&A...402..211H, Meixner2004ApJ...614..371M} are
respectively the masses of CO and amorphous carbon in the HD\,56126
envelope, $\mu_{\rm CO}$ and $\mu_{\rm C}$ are the molecular
(atomic) weights of CO and C, respectively.
We estimate the total hydrogen mass to be
\begin{equation}\label{eq:mH}
M_{\rm H} = \frac{1.4\,M_{\rm CO}/\mu_{\rm CO}
                  + M_{\rm AC}/\mu_{C}}
                 {\CtoH_\star}
\approx 0.20\,M_\odot ~~.
\end{equation}
This corresponds to a total envelope mass $M_{\rm env}\approx
0.28\,M_\odot$ and a gas-to-dust ratio of $\rgd\approx 380$. With
all the sulfur atoms in MgS (i.e. $\Smgs$\,=\,1), we obtain $M_{\rm
MgS}^{\rm ava} \approx 4.56\times 10^{-5}\,M_\odot$. This is much
smaller than $M_{\rm MgS}^{\rm req}$, the amount of MgS mass
required to explain the power emitted at the 30$\mum$ feature (see
Table\,1).

We should note that the actual available MgS mass could be even
lower since it is likely that some S atoms are tied up in gas
molecules such as SiS and CS. In the C-rich AGB star CW Leo
(IRC\,+10216), SiS and CS consume a substantial fraction of the S
atoms \citep{Glassgold1996ARA&A..34..241G,
Millar2001MNRAS.327.1173M}. If this also holds for HD\,56126, the
shortage of MgS dust mass would become more severe.
It does not help if MgS mixes with amorphous carbon.\footnote{%
  Goebel \& Moseley (1985) argued that MgS could form via chemical
  surface reactions on carbon grains.
  }
As illustrated in Figure\,\ref{fig:1}, $Q_{\rm o}$\,=\,1.6 and
$\lambda_{\rm o}$\,=\,$\pi a$ (adopted for MgS) also overestimate
the UV/visible absorptivity of amorphous carbon. The mixture of
amorphous carbon and MgS would have a lower temperature since the IR
emissivity of amorphous carbon is much larger than that of MgS.
Therefore one would expect a larger $M_{\rm MgS}^{\rm req}$.
We should also note that HD\,56126 is not the strongest
30$\mum$ feature source. The MgS budget problem would be
more severe for objects with a stronger 30$\mum$ band
and cooler MgS dust (e.g. IRAS\,19454, a post-AGB star,
requires $T\approx 50\K$; see \citealt{Hony2002A&A...390..533H}).

\citet{Zhukovska2008A&A...486..229Z} recently performed
a detailed study of the condensation of MgS in the outflows
from C-rich stars on the tip of the AGB. They found that
MgS can only be formed by precipitating on pre-existing
grains and therefore one would expect MgS to form as a mantle
on the SiC core. However, these grains would exhibit a feature
at $\simali$33--38$\mum$ which is not seen in C-rich objects.

\citet{Hony...2003A&A...402..211H} and \citet{HonyBouwman2004}
assumed an analytic formula for
the UV/visible absorption efficiency of MgS dust,
which differs from eq.1 in that theirs have an
appreciable absorbing power at $0.3<\lambda<2\mum$:
\begin{equation}\label{eq:qabs_UV_Hony}
\qabsuv(\lambda) = \left\{\begin{array}{lr}
1 ~, ~~~~~~~~ \lambda \le 1\mum ~;\\
2 - \lambda ~, ~~ 1 < \lambda \le 2\mum ~;\\
0 ~, ~~ \lambda > 2\mum ~.\\
\end{array}\right.
\end{equation}
With this kind of $\qabsuv(\lambda)$ and assuming a spatial
distribution of $dn/dr \propto r^{-1}$ (within a narrow
zone of 1.2$^{\prime\prime}$--2.6$^{\prime\prime}$ from the star)
for the dust and a gas-to-dust ratio of $r_{\rm g/d} = 600$,
\citet{Hony...2003A&A...402..211H} performed detailed radiative transfer
modeling calculations for HD\,56126 and found that the S abundance
constraint was not violated.

If we adopt the UV/visible absorption efficiency $\qabsuv(\lambda)$
of \citet{Hony...2003A&A...402..211H}, for MgS dust of $a=0.1\mum$
we would need $M_{\rm MgS}^{\rm req} \approx 8.38\times
10^{-4}\,M_\odot$ and $4.56\times 10^{-4}\,M_\odot$ respectively for
the dust spatial distributions of $dn/dr \propto r^{-1}$ and
$dn/dr \propto r^{-2}$, far exceeding the maximum amount of MgS dust
$M_{\rm MgS}^{\rm ava}$ that could be available in HD\,56126
(see Table 2).\footnote{%
    Following Meixner et al.\ (2004), we assume
    the dust to be distributed in a broader zone, i.e.,
    1.2$^{\prime\prime}$--7$^{\prime\prime}$ from the star.
    }
For $a=0.01\mum$, the required MgS mass is reduced to
$M_{\rm MgS}^{\rm req} \approx 8.01\times 10^{-5}\,M_\odot$
and $4.69\times 10^{-5}\,M_\odot$ respectively for
$dn/dr \propto r^{-1}$ and $dn/dr \propto r^{-2}$,
and appears comparable to $M_{\rm MgS}^{\rm ava}$.
However, we note that (1) $M_{\rm MgS}^{\rm ava}$ is already
an upper limit, and (2) we are not sure if eq.11 is a reasonable
approximation for the UV/visible absorption properties of MgS dust
of $a=0.01\mum$. As shown in Figure 1, the UV/visible absorption
cut-off wavelength $\lambdao$ varies with dust size $a$.
For the dust species listed in Figure 1, eq.1 seems to be
a more reasonable approximation than eq.11
for their $\qabsuv(\lambda)$.
We call on laboratory measurements of the UV/visible/near-IR
dielectric functions of MgS dust.

Finally, we calculate $E_{\rm abs}^{\rm tot}$ ($\erg\s^{-1}\g^{-1}$)
--- the total power absorbed by
one gram MgS dust of size $a$ from
\begin{equation}\label{eq:Eabs1}
E_{\rm abs}^{\rm tot} = \frac{\rstar^2}{2}
\frac{\ln\left(r_{\rm out}/r_{\rm in}\right)}
{r_{\rm out}^2-r_{\rm in}^2}
\int_{912\Angstrom}^{\infty}
\left[3\,Q_{\rm abs}(a,\lambda)/4a\rho_{\rm MgS}\right]\,\Fstar\,d\lambda
~~~,~~{\rm for}~~ dn/dr \propto r^{-1} ~~~;
\end{equation}
\begin{equation}\label{eq:Eabs2}
E_{\rm abs}^{\rm tot} = \frac{\rstar^2}{4}
\frac{1}{r_{\rm in}\,r_{\rm out}}
\int_{912\Angstrom}^{\infty}
\left[3\,Q_{\rm abs}(a,\lambda)/4a\rho_{\rm MgS}\right]\,\Fstar\,d\lambda
~~~,~~{\rm for}~~ dn/dr \propto r^{-2} ~~~.
\end{equation}
If we assume the absorbed energy is {\it all} radiated
away through the 30$\mum$ feature, we would require
a total MgS mass of $M_{\rm MgS}^{\rm min} =
E_{30\mu{\rm m}}^{\rm obs}/E_{\rm abs}^{\rm tot}$
--- apparently, this is absolutely a lower limit
since the absorbed energy will not be {\it exclusively}
emitted from the 30$\mum$ feature (a fraction of the energy
will be radiated away at other wavelengths and through
the continuum underneath the 30$\mum$ feature).
In Table 3 we tabulate $E_{\rm abs}^{\rm tot}$
and  $M_{\rm MgS}^{\rm min}$.
We see in all cases $M_{\rm MgS}^{\rm min} > M_{\rm MgS}^{\rm ava}$,
indicating that we simply do not have enough MgS to account
for the observed emission power.

\section{Summary}\label{sec:summary}
We have investigated the hypothesis of MgS as a
carrier of the prominent 30$\mum$ emission feature seen in numerous
C-rich evolved objects, using HD\,56126 as a test case. It is found
that, in order to account for the enormous power emitted from this
feature, one requires a much higher MgS dust mass than available in
this object. We therefore argue that MgS is unlikely the carrier of
the 30$\mum$ feature.

\acknowledgments{We thank S. Hony and A.K. Speck for
very helpful discussions or comments.
BWJ and KZ are supported by China 973 Program
No.\,2007CB815406. AL is supported in part by Spitzer Theory
Programs and NSF grant AST 07-07866.}

%


\clearpage

\begin{table}
\caption{\label{tab:mgsmass_1}
         $M_{\rm MgS}^{\rm req}$
         (in unit of 10$^{-5}$\,$M_{\odot}$)
         --- The MgS mass required to account for
         the power emitted in the 30$\mum$ feature
         in HD\,56126.
         }
\begin{center}
\begin{tabular}{ccc}
\hline
\hline
\multicolumn{3}{c}{spherical dust (Mie theory)}\\
\hline
Dust Size & $dn/dr$  & $dn/dr$\\
($\mum$)  & $\varpropto r^{-2}$ &$\varpropto r^{-1}$\\
\hline
0.01 & 46.1 &84.9 \\
0.05 & 46.1 &84.9\\
0.1  & 47.5 &87.5 \\
0.3  & 87.3 &164 \\
\hline
\multicolumn{3}{c}{CDE}\\
\hline
0.01 & 34.7 &64.3 \\
0.05 & 34.7 &64.3\\
0.1  & 35.7 &64.3 \\
0.3  & 65.9 &126 \\
\hline
\end{tabular}
\end{center}
\end{table}

\begin{table}
\caption{\label{tab:mgsmass_2}
         $M_{\rm MgS}^{\rm ava}$
         (in unit of 10$^{-5}$\,$M_{\odot}$)
         --- The MgS mass that could be available in
         the envelope of HD\,56126, with the assumption
         that all S atoms are tied up in MgS.
         }
\begin{center}
\begin{tabular}{ccccr}
\hline
\hline
$M_{\rm dust}$ & gas/dust & $M_{\rm env}$
               & $M_{\rm MgS}^{\rm ava}$ & References\\
($10^{-4}\,M_\odot$) & ($\rgd$) & ($M_\odot$)
               & ($10^{-5}\,M_\odot$) & \\
\hline
7.8 & 75 & 0.059
         & 0.96 & Meixner et al.\ (2004)\\
7.4 & 220 & 0.16
          & 2.60 & Hony et al.\ (2003)\\
7.4 & 600 & 0.44
          & 7.16 & Hony et al.\ (2003)\\
7.4 & 380 & 0.28
          & 4.56 & this work\\
\hline
\end{tabular}
\end{center}
\end{table}

\begin{table}
\caption{\label{tab:mgsmass_3}
         $M_{\rm MgS}^{\rm min}$ --- The minimum amount of
         MgS dust required. This is obtained by assuming
         that the energy absorbed by a MgS grain is exclusively
         emitted through the 30$\mum$ band (see \S6).
         Also tabulated is $E_{\rm abs}^{\rm tot}$ ($\erg\s^{-1}\g^{-1}$),
         the power per unit mass absorbed by MgS dust.
         We see in all cases
         $M_{\rm MgS}^{\rm min} > M_{\rm MgS}^{\rm ava}$.
         }
\begin{center}
\begin{tabular}{ccccc}
\hline Dust Size &
\multicolumn{2}{c}{$dn/dr\varpropto r^{-2}$}
            & \multicolumn{2}{c}{$dn/dr\varpropto r^{-1}$}\\
\cline{2-5} ($\mum$) & $E_{\rm abs}^{\rm tot}$ & $M_{\rm MgS}^{\rm min}$
                     & $E_{\rm abs}^{\rm tot}$ & $M_{\rm MgS}^{\rm min}$\\
 $a$ & ($\erg\s^{-1}\g^{-1}$) & ($M_\odot$)
     & ($\erg\s^{-1}\g^{-1}$) & ($M_\odot$)\\
\hline
0.01 & $2.85\times10^{6}$ & $3.53\times10^{-4}$
     & $1.62\times10^{6}$ & $6.20\times10^{-4}$ \\
0.05 & $2.85\times10^{6}$ & $3.53\times10^{-4}$
     & $1.62\times10^{6}$ & $6.20\times10^{-4}$\\
0.1  & $2.77\times10^{6}$ & $3.62\times 10^{-4}$
     & $1.58\times10^{6}$ & $6.38\times10^{-4}$ \\
0.3  & $1.57\times10^{6}$ & $6.42\times10^{-4}$
     & $8.90\times10^{5}$ & $1.13\times10^{-3}$ \\
\hline
\end{tabular}
\end{center}
\end{table}

\end{document}